\documentstyle[11pt]{article}
\begin{document}
\makeatletter \@addtoreset{equation}{section} \makeatother
\renewcommand{\theequation}{\thesection.\arabic{equation}}
\baselineskip 15pt

\title{\bf Book Review: \\ ``Quantum Theory as an Emergent Phenomenon'' \\
by Stephen L. Adler \\ {\small\bf Cambridge University Press
(2004)}}
\author{Angelo Bassi\footnote{e-mail: bassi@mathematik.uni-muenchen.de}\\
{\small Mathematisches Institut der Universit\"at M\"unchen,} \\
{\small Theresienstr. 39, 80333 M\"unchen, Germany, and}\\
{\small The Abdus Salam International Centre for Theoretical
Physics,} \\
{\small Strada Costiera 11, 34014 Trieste, Italy.}}
\date{}
\maketitle

Stephen Adler's new book, ``Quantum Theory as an Emergent
Phenomenon'', is intriguing already from the title: why should one
consider quantum mechanics, our most successful physical theory
which has been confirmed by all experiments performed up to now,
not as a fundamental theory of nature but rather as an effective
theory emerging from a deeper level of dynamics? Of course, here
we are not referring just to a specific quantum mechanical model
but to the general mathematical structure of quantum theory, where
states of physical systems are represented by vectors of suitable
Hilbert spaces, observables by linear operators and the dynamics
by a linear equation for the vectors.

Indeed, there is a deep fundamental reason why quantum theory as
it stands can not be our ultimate theory, which goes under the
name of the {\it measurement problem}. Due to the {\it linearity}
of quantum mechanics, microscopic systems can be easily prepared
in a superposition of two different states, and when they interact
with macroscopic objects, as it happens e.g. in a measurement
process, they easily generate a superposition of two
macroscopically different states of the macro-object. The question
is: why do we not see such kind of macroscopic superpositions? Why
do macroscopic objects seem to be always well localized in space?
The standard answer to these questions is provided by the
Copenhagen interpretation, according to which when a micro-system
interacts with a macro-object and creates a superposition of
different macroscopic states, the unitary evolution given by the
Schr\"odinger equation is immediately followed by a reduction
process which randomly drives the superposition into one of its
terms, with a probability distribution given by the Born rule.
This is the so called {\it postulate of wavepacket reduction}
which should explain both the fact that measurements on quantum
systems always have definite outcomes and the probability of such
outcomes.

Thus, the Copenhagen interpretation is in deep conflict with our
idea that a unique type of evolution should at least in principle
rule all types of physical processes; worse than that, it does not
tell us in a clear, precise and unambiguous way under which
circumstances a system evolves according only to the Schr\"odinger
equation and in which other cases also wavepacket reduction
occurs. The problem of understanding quantum theory is then still
open.

The measurement problem is the starting point of Adler's book, and
the solution he proposes is the following: {\it quantum mechanics}
is not a fundamental theory of nature but {\it an emergent
phenomenon arising from the statistical mechanics of matrix models
with a global unitary invariance}. The book is entirely devoted to
showing how that idea can be implemented within a concrete (and
highly sophisticated) mathematical framework.

The first chapters of the book describe the mathematical and
physical structure of the so called {\it trace dynamics}: the
dynamical variables of the theory are noncommuting matrices
$q_{r}(t) \equiv [q_{r}(t)]_{ij}$, where $i$ and $j$ denote the
matrix elements
--- e.g. the analog of the classical Klein-Gordon scalar field
$\phi(x,t)$ is a matrix field $[{\phi}(x,t)]_{ij}$
--- whose dynamics is formulated in terms of a trace Lagrangian and
trace Hamiltonian generalizing the classical Lagrangian and
Hamiltonian formalism to a phase space of noncommuting variables;
contrary to quantum mechanics, no a priori commutation relations
between the matrices are assumed. The relevant mathematical tools
and properties of trace dynamics are described: cyclic identities,
operator derivatives, generalized Poisson brackets, conserved
quantities; the extension to supersymmetric theories is discussed.

After these mathematical preliminaries, the discussion gets
directly to the core. The statistical mechanics of the trace
dynamics is set up: the analog of the classical Liouville theorem
is proven, the canonical ensemble is analyzed, showing how average
quantities can be computed. The author next shows how under
plausible assumptions an effective theory stems from the
statistical mechanics of the canonical ensemble, which is
equivalent to quantum mechanics: the key idea is that global
unitary invariance implies the existence of a conserved Noether
charge, the equipartition of which gives rise to the canonical
algebra of quantum mechanics. More specifically, the effective
variables-matrices $q_{r \makebox{\tiny eff}}(t)$ and effective
conjugate momenta-matrices $p_{r \makebox{\tiny eff}}(t)$ of the
emerging theory are shown to evolve under a unitary dynamics and
to satisfy the quantum commutation relations, with the constant
appearing in the commutation relations identified with $\hbar$.
This is the crucial claim of the entire book: quantization is {\it
not} obtained in the standard way by promoting each degree of
freedom $[q_{r}(t)]_{ij}$ and its conjugate momentum
$[p_{r}(t)]_{ij}$ to quantum operators satisfying specific
commutation relations. On the contrary, quantization naturally
{\it emerges} from the statistical mechanics of the underlying
dynamics: the quantum-like dynamical variables $q_{r
\makebox{\tiny eff}}(t)$ and $p_{r \makebox{\tiny eff}}(t)$ are
the result of thermodynamic averages over functions of the true
variables $q_{r}(t)$ and $p_{r}(t)$.

And there is still something more. In the final part of the book
the author considers the effects of statistical fluctuations
around the averages and the result he gets is quite surprising.
While thermodynamic averages give rise to the quantum framework,
i.e. basically to the usual Schr\"odinger equation, fluctuations
around the averages show up as a stochastic modification of the
Schr\"odinger dynamics precisely of the type considered within the
context of spontaneous collapse models. According to these models
there is a unique dynamical equation both for microscopic and
macroscopic systems: in the microscopic case, the dynamics
practically reduces to standard quantum mechanics, while in the
macroscopic case it reproduces classical mechanics and in
particular predicts that in measurement-like situations
wavefunctions of macroscopic measuring systems do not remain in a
superposition of different states corresponding to the different
outcomes, but spontaneously reduce to just one state, with a
probability which is almost identical to the Born probability
rule, as predicted by the postulate of wavepacket reduction. It is
important to stress that here the reduction process is not just an
ad hoc mechanism which is put in by hand to avoid embarrassing
macroscopic superpositions, but is embodied in the modified
Schr\"odinger equation. Note also that the author provides an
explanation for the origin of the noisy terms which are
responsible for the reduction process.

The goal of the entire book is now reached and the emerging
physical picture is the following: physical systems
--- this is the ontology suggested by the theory  --- are
represented by matrices with no a priori commutativity properties,
whose dynamics, which in general is neither unitary nor local,
should describe all physical phenomena. The quantum mechanical
behavior of microscopic systems is recovered as a result of the
thermodynamical analysis of the fundamental dynamics; the emerging
quantum dynamics is governed by a modified stochastic
Schr\"odinger equation which reproduces the standard Schr\"odinger
evolution at the microscopic level, but at the same time implies
that macroscopic objects are described by wavefunctions which are
localized in space and obey the classical laws of motion, avoiding
in this way the measurement problem of quantum mechanics.

Needless to say, there is still much work to be done and technical
and general questions to answer: in particular, a candidate model
fulfilling all the assumptions which lead to the emergence of
quantum theory out of the statistical mechanics of trace dynamics
has not been identified yet; the connection between the formalism
and our perceptions about physical objects needs to be analyzed in
greater detail.

To conclude, the book describes in detail a rich mathematical
framework out of which a new physical description of nature can
emerge. The chapters are well organized; the presentation is
clear, rigorous, very attractive, and accessible to mathematical
as well as theoretical physicists. It is recommended to everyone
who wants to learn more about the fascinating field of the
foundations of physics and the continuing efforts scientists are
making to formulate a clear and coherent picture of physical
phenomena.
\end{document}